# AI-based System for Transforming Text and Sound to Educational Videos

**M. E. ElAlami[1,*], S. M. Khater[2], M. El. R. Rehan[3]**

[1]Prof. of Computer and Information System, Faculty of Specific Education, Mansoura University, Egypt

[2]Lecturer of computer teacher preparation Department, Faculty of Specific Education, Mansoura University, Egypt

[3]Demonstrator of computer teacher preparation Department, Faculty of Specific Education, Mansoura University, Egypt

Emails: moh_elalmi@mans.edu.eg; shim_khater@mans.edu.eg; mohamedrehan@mans.edu.eg

**Abstract**

Technological developments have produced methods that can generate educational videos from input text or sound. Recently, the use of deep learning techniques for image and video generation has been widely explored, particularly in education. However, generating video content from conditional inputs such as text or speech remains a challenging area. In this paper, we introduce a novel method to the educational structure, Generative Adversarial Network (GAN), which develop frame-for-frame frameworks and are able to create full educational videos. The proposed system is structured into three main phases in the first phase; the input (either text or speech) is transcribed using speech recognition. In the second phase, key terms are extracted and relevant images are generated using advanced models such as CLIP and diffusion models to enhance visual quality and semantic alignment. In the final phase, the generated images are synthesized into a video format, integrated with either pre-recorded or synthesized sound, resulting in a fully interactive educational video. The proposed system is compared with other systems such as TGAN, MoCoGAN, and TGANS-C, achieving a Fréchet Inception Distance (FID) score of 28.75%, which indicates improved visual quality and better over existing methods.

## 1. Introduction

Visual media especially video plays a central role in modern education by enhancing learner engagement and comprehension across disciplines. Studies confirm that well-crafted educational videos can simplify complex topics, support retention, and address varied learning styles [1]. Simultaneously, Artificial Intelligence (AI) is revolutionizing education by enabling personalized learning, automating instructional tasks, and offering analytical insights. AI's data processing capabilities contribute to more adaptive and effective educational practices [2]. The rise of Multimodal Learning which involves combining text, audio, images, and video further enriches educational experiences. AI-powered Multimodal Learning Analytics (MMLA) allow deeper understanding of learner interactions, supporting personalized feedback and learning design. [3].

Generative AI, particularly Multimodal Large Language Models (MLLMs), now enables the creation of diverse educational content, including dynamic. These models offer scalable, personalized, and engaging resources, significantly advancing multimodal teaching approaches [4]. Despite the capabilities of Generative AI, producing high-quality educational videos that seamlessly integrate text and dynamic visuals remains technically demanding and time-consuming. Traditional production methods require significant expertise and resources, limiting their widespread use in educational contexts [5]. To address the challenges of manual production and leverage the capabilities of Generative AI, there is growing interest in developing intelligent systems capable of automating the educational video content creation process [6].

This paper aims to design and develop an innovative intelligent system capable of automatically converting textual and auditory inputs into integrated educational videos. The primary contribution of this study lies in presenting detailed system architecture and intelligent algorithms that address the challenges of video generation from multimodal sources (text and sound), focusing on producing pedagogically coherent and visually engaging content. The research addresses the current gap in automated educational video creation tools by proposing a solution that facilitates the efficient and effective production of high-quality visual learning materials for educators and developers, thereby contributing to the enhancement and scalability of video use in education.

## 2. Related work

Yan et al (2024) [7] introduced a novel approach to time representation in video synthesis, where a continuous video generator based on neural representations was developed. Motion representations were constructed using positional embedding, and sparse training strategies for video generators were explored. They redesigned a dual-structure video discriminator. StyleGAN2 served as the foundation for the model, which has benefits including effective training, excellent image quality, and an adjustable latent space. It was suggested that this work provides a strong foundation for future advancements in video generation.

JayZhangjieWu et al (2024) [8] introduced a singular assessment metric for evaluating text-to-video technology. It consists of two additives: T2VScore-A for measuring alignment between video and text, and T2VScore-Q for comparing the visual quality based on technical and semantic metrics. The authors also created the TVGE dataset, containing over 2,500 human-rated text-to-video samples. Experimental results show that T2VScore significantly outperforms existing metrics in terms of correlation with human judgment, offering a more reliable and comprehensive method for evaluating AI-generated videos.

Mao et al. (2024) [9] introduced a new system known as generation of audible videos from text (TAVG), which encompasses the creation of simultaneous video and audio based on text-only descriptions. To achieve this, researchers delivered a new repertoire of more than 1.7 million videos with accurate automatic audio and image descriptions using ChatGPT tools and enhancements. They also proposed a new metric to measure audio-video compatibility. To provide a standard model, they presented a generative model that uses latent diffusion with contrastive learning and cross-attention techniques to ensure synchronization between outputs. Experiments demonstrated the superiority of this model over traditional approaches in terms of content quality and consistency, opening new avenues in multimodal generation.

Yue Ma, et al. (2023) [10] had presented Pose-Guided Text-to-Video Generation using Pose-Free Videos, which addresses the growing demand for generating text-editable and pose-controllable character videos a crucial component in creating diverse digital humans. This task has long been limited by the lack of comprehensive datasets containing paired video-pose captions and the absence of suitable generative prior models for video synthesis. To overcome these limitations, they proposed a novel two-stage training scheme that leverages easily obtainable datasets such as image pose pairs and pose-free videos in conjunction with a pre-trained Text-to-Image (T2I) model, enabling effective generation of pose-controllable character videos.

Zhu et al (2023) [11] had presented Text-Image-to-Video Generation: Controllable Video Synthesis from Static Images and Text Descriptions, which addresses the challenge of generating controllable videos that align with user intentions a highly appealing yet complex task in computer vision. They proposed a unique job dubbed Text-Image-to-Video generation (TI2V) to allow for fine-grained control over both appearance and motion. This method enables the generation of videos from a combination of a static image and a text description, providing the flexibility to control both visual elements and dynamic motion in the synthesized video, meeting the user's specific requirements.

Singer et al (2022) [12] presented Make-A-Video, a system that extends Text-to-Image (T2I) generation to Text-to-Video (T2V). It learns visual appearance from text-image pairs and motion from unsupervised video data. The model speeds up training, avoids the need for paired text-video data, and inherits diverse image generation capabilities. Spatial-temporal modules, including a decomposed temporal U-Net and attention tensors, are used. A spatial-temporal pipeline generates high-resolution, high-frame-rate videos. The approach sets a new state-of-the-art in text-to-video generation, confirmed by both qualitative and quantitative results.

After reviewing previous relevant research, several limitations were identified in existing systems, including the lack of integration between text input and audio output, limited capability in handling complex educational content, generation of static or short-duration videos with poor semantic alignment, and suboptimal video quality. To overcome these challenges, a novel system Text to Video Adversarial Video Generation Network is proposed. This system extends traditional text-to-video adversarial models by enabling the generation of animated educational videos from textual input. It employs advanced techniques such as text encoding and keyword extraction using the YAKE algorithm, while also linking visual elements to appropriate audio segments from a predefined sound database. The goal is to produce semantically rich, coherent, and high-quality educational videos that align closely with the input text.

### 3. The Proposed System

This study used deep learning to convert text or sound into educational videos. The system consists of three stages. In the first stage, the user provides an input sentence, either in the form of text or audio. The system then converts sound to text. Using the Bidirectional Encoder Representations from Transformers (BERT) model, important and domain-specific terms are extracted from the input. For each extracted term, corresponding images are generated from an image database using GAN technique. In the second stage, the selected images for each term are combined according to the user's arrangement to create a video without sound composed of the selected images. In the third stage, the appropriate audio files are selected for each term, after which the audio file is merged with the video. In the end, the video is displayed and can be downloaded to the user's device. In the following paragraph, the proposed will be explained in detail.

The user enters basic words or sentences through the text box or through the microphone. Speech is converted into text using Google Assistant. The speech then appears as text in the text box. After that, the system searches for images related to the sentence or word in the image dataset. Each keyword has a set of stored images. Images are generated using a Vector Quantum Generative Adversarial Network (VQGAN). This process helps to create images with more defined structures and sharp edges compared to traditional GANs. The CLIP model is used to match image and text based on similarity and semantic alignment. Diffusion models are also used. They help enhance image quality by denoising, refining details, and applying visual filters. The processed images are compiled and rendered into a coherent video sequence. Sound tracks are then integrated into the video. This ensures an interactive and immersive multimedia experience. The framework of the proposed system as shown in figure (1).

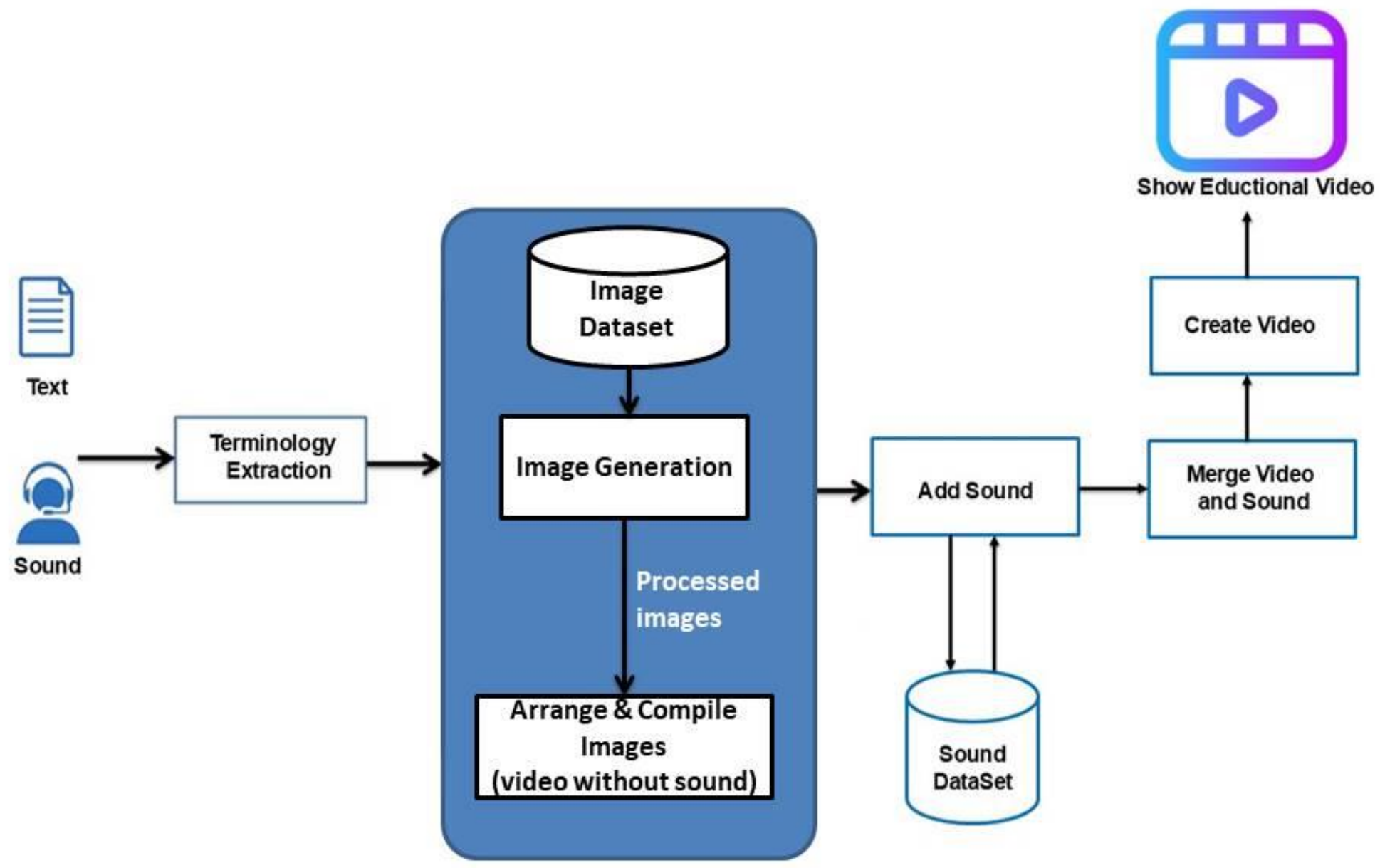

**Figure 1.** Framework of the proposed system

The ability of GANs to produce crisp images has been demonstrated. From this perspective, we may be able to generate videos more efficiently if we begin our text-to-video network training with a text-to-image stage. We divided the training procedure into two phases, Text-to-Image Generation and Evolutionary Generation, based on these concepts. We describe the general flow of our suggested architecture. We start by learning how to generate text-to-single images, and then progressively increase the quantity of photos that are generated. This training process is repeated until the desired video duration is reached. This is the main paradigm we use. The methods we employed to stabilize the learning will be explained once these two phases are thoroughly explained in subsections. Figure (2) explain flow chart of the proposed system.

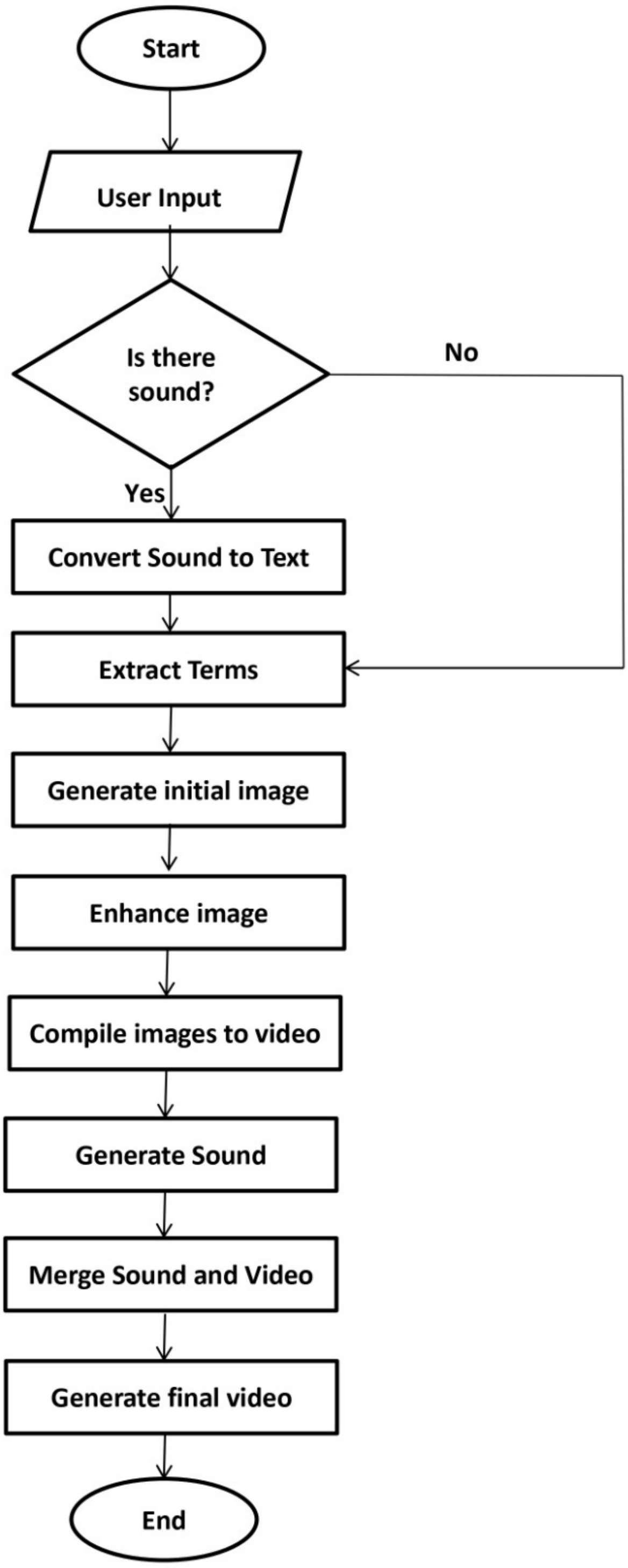

**Figure 2.** Flowchart of the proposed system

The following algorithm outlines the systematic steps of the proposed Educational Video Generation System. It integrates advanced AI models such as BERT for text analysis, VQGAN and Diffusion for image generation, and CLIP for semantic alignment. The goal is to automatically transform user-inputted text or audio into engaging, semantically accurate educational videos. Table 1 presents the systematic workflow of the system.

**Table 1**: The Proposed Algorithm for Educational Video Generation System

| Step No. | Step Function |
|---|---|
| **Step 1** | **Initialization**<br>Initialize GUI with input textbox and microphone button.<br>Load pre-trained models: BERT, VQGAN, CLIP, Diffusion.<br>Prepare storage paths for image and video data. |
| **Step 2** | **Input Acquisition**<br>If the user inputs audio, convert it to text using Google Assistant or Whisper.<br>If the user inputs text, pass it directly to the system.<br>Display the resulting text in the input box. |
| **Step 3** | **Keyword Extraction**<br>Use BERT to extract key domain-specific terms from the input text.<br>Store all keywords in a list for further processing. |
| **Step 4** | **Image Generation•**<br>**For each keyword:**<br>Generate image using VQGAN model.<br>Use CLIP model to match generated image with text semantically.<br>Apply Diffusion model to enhance image quality.<br>Store final image for each term in order. |
| **Step 5** | **Video Composition**<br>Arrange selected images sequentially based on keyword order or user input.<br>Compile the images into a silent video stream. |
| **Step 6** | **Audio Integration**<br>Retrieve or generate audio tracks corresponding to the extracted terms.<br>Merge audio with the video to create a final educational video. |
| **Step 7** | **Video Output**<br>Display the final video on the GUI.<br>Provide a button to allow user to download the generated video.<br>End of process. |

**Dataset Source:**

The proposed machine uses publicly to be had datasets. For picture generation, we make use of datasets from Kaggle, along with Common Objects in Context (COCO), which provides classified photos for education. For speech processing, we appoint the LibriSpeech dataset (available on Kaggle) containing 1,000 hours of English audio.

**Algorithm 1: Training Procedures**

**(I)Data:**

• Input text: t

• Corresponding real video frames: $V_R = (X_1, X_2, ..., X_\{2^n\})$

Network Components:

• Generator: G

• Gated Recurrent Unit (GRU): R

• Discriminators: $D_I$, $D_{S_1}$, $D_{S_2}$, ..., $D_{S_n}$

**(II) Stage of Text-to-Image Generation**

Objective: To generate a single image from text input and train the networks G, R, and the image discriminator $D_I$.

while not converged do

Get $z_0, z_1 \in (0, 1)$

Generate $I_1 = G(R(z_0, (\varphi_t, z_1)))$

Randomly choose one real image X from $V_R$

Update G, R to minimize loss according to Equation 3

Update $D_I$ to maximize loss according to Equation 3 using independent samples pairing

end

**(III) Stage of Evolutionary Generation**

Objective: To iteratively generate sequences of images and refine both spatial and temporal consistency using an evolutionary process.

for m ⇒ 1 to n do

while not converged do

Get $z_1, ..., z_{2^m} \in \mathcal{N}(0, 1)$

Generate $I_1, ..., I_{2^m}$ by repeating R and G

Randomly choose $2^m$ consecutive real frames $X_1, ..., X_{2^m}$ from $V_R$

Update G, R to minimize loss according:

- Loss function for image-level discrimination
- Loss function for temporal sequence-level discrimination

Update $D_I$ to maximize loss via Equation 3 using independent samples pairing

Update $D_{S_m}$ to maximize loss via Equation 4

end

end

**Where:**

- t : Input text
- $V_R$ : Sequence of real video frames
- $X_i$ : The i-th frame in the real video
- G : Generator network
- R : GRU network for processing encoded text
- $D_I$ : Discriminator for individual image quality
- $D_{S_m}$ : Discriminator for temporal sequence consistency at level m
- z : Noise vector sampled from normal distribution $\mathcal{N}(0,1)$
- $\varphi_t$ : Encoded representation of text t
- $I_i$ : Generated image(s)

**Training Procedures**

First, the new step-discriminator $DS_1$ is initialized using D1, the image discriminator. All step-discriminators are required to have identical architectures, except for the number of input channels in their first layer. Specifically, $DS_m$ is provided with $2^m$ images as input, while $DS_{m-1}$ is given $2^{m-1}$ images. The initialization of these step-discriminators can thus be described as follows:

Strikingly wrong training architecture of both generators and the discriminator is used in our practice. However, unlike a normal GAN, a small perturbation takes place in the order between two branches in the discriminator. Here, the discriminator goes into several conversion teams to achieve a high level of function mapping. One branch, with Sankranti on Sankranti, computes the texture wallpaper matching of the text image, while the other branch discriminates without a text solver. The activities of these two separate branches enhance the image's quality and accurately match the image to the lesson. The model for lesson matching is trained with three types of loss, similarly to Reed et al. Actual pairs (x, φt), fake pairs (I, φt), and wrong pairs (x, t), with φt being a specific text vector that is different from φt.

The penalties generated by G,R, D, and D_S are general descriptions of the following:

$$L_{obj}(G, R, D_I, D_S) = L_I(G, R, D_I) + L_S(G, R, D_S) \quad \textbf{(1)}$$

This equation combines two components:

- $L_I(G, R, D_I)$: the adversarial loss for individual image-level discrimination.

- $L_S(G, R, D_S)$: the adversarial loss for sequence-level or video segment discrimination.

Image-Level Loss: $L_I(G, R, D_I)$

$$L_I(G, R, D_I) = \Sigma_i [ \log(D_I(X^i)) + \log(D_I(X^i, \varphi_i)) + \log(1 - D_I(I^i)) + \log(1 - D_I(I^i, \varphi_i)) + \log(1 - D_I(X^i, \hat{\varphi}_i)) ] \quad \textbf{(2)}$$

This function measures how well the discriminator D_I distinguishes real images and their correct descriptions from generated or mismatched ones.

**Where:**

- $X^i$: a real image.

- $I^i$: a generated image.

- $\varphi_i$: matching text or conditioning feature.

- $\hat{\varphi}_i$: mismatched or irrelevant feature.

- $D_I$: the image discriminator.

**Sequence-Level Loss: $L_S(G, R, D_S)$**

$$L_S(G, R, D_S) = \Sigma_i [ \log(D_S(X^{i,S})) + \log(D_S(X^{i,S}, \varphi_i)) + \log(1 - D_S(I^{i,S})) + \log(1 - D_S(I^{i,S}, \varphi_i)) + \log(1 - D_S(X^{i,S}, \hat{\varphi}_i)) ] \quad \textbf{(3)}$$

**Where:**

- $X^{i,S}$: a real video/image sequence.

- $I^{i,S}$: a generated video/image sequence.

- $D_S$: the sequence (or video) discriminator.

This loss function ensures that not only individual frames are realistic, but the temporal consistency and semantic alignment across sequences are preserved.

**4. Applications and Results**

The proposed system was designed using HTML, CSS, PHP, MYSQL and Node.js. The GAN algorithm and TiVGAN method are used to create tutorial videos from text or sound inputs. The system also has a separate image database and database of sound files to improve the generation process by providing relevant visual and sound content and through processes within the system. The images are extracted from the database based on the terms extracted. Each term has a collection of images. Then the images are selected from among the images displayed for each term.

The video is generated from these selected images. The video is then integrated with the selected sound files from the sound database. The user selects each term of the files and then the full video is generated at the end. Running the proposed system on a computer requires the availability of XAMPP software to convert it to a local server, as well as some Node.js libraries.

The proposed system was applied to configured computing devices with 16 GB RAM, an Intel Core i7-6820HQ processor working on the frequency of 2.70 GHz, and a dedicated graphics card with 4 GB memory. System 64-bit operating system operates on architecture. These specifications were sufficient to efficiently execute computational functions involved in video generation, image processing and multimedia synchronization within the proposed structure

The figure 3 shows where the user inserts text or audio. When pressing the **"Extract Terms from Inserted Sentence"** button, the terms are displayed in a table. The first column contains the term, the second column shows the corresponding audio file, and the third column displays images for each term. After that, the user selects the desired images from those displayed using TiVGAN technology. When the **"Create Video"** button is pressed, the video creation process begins.

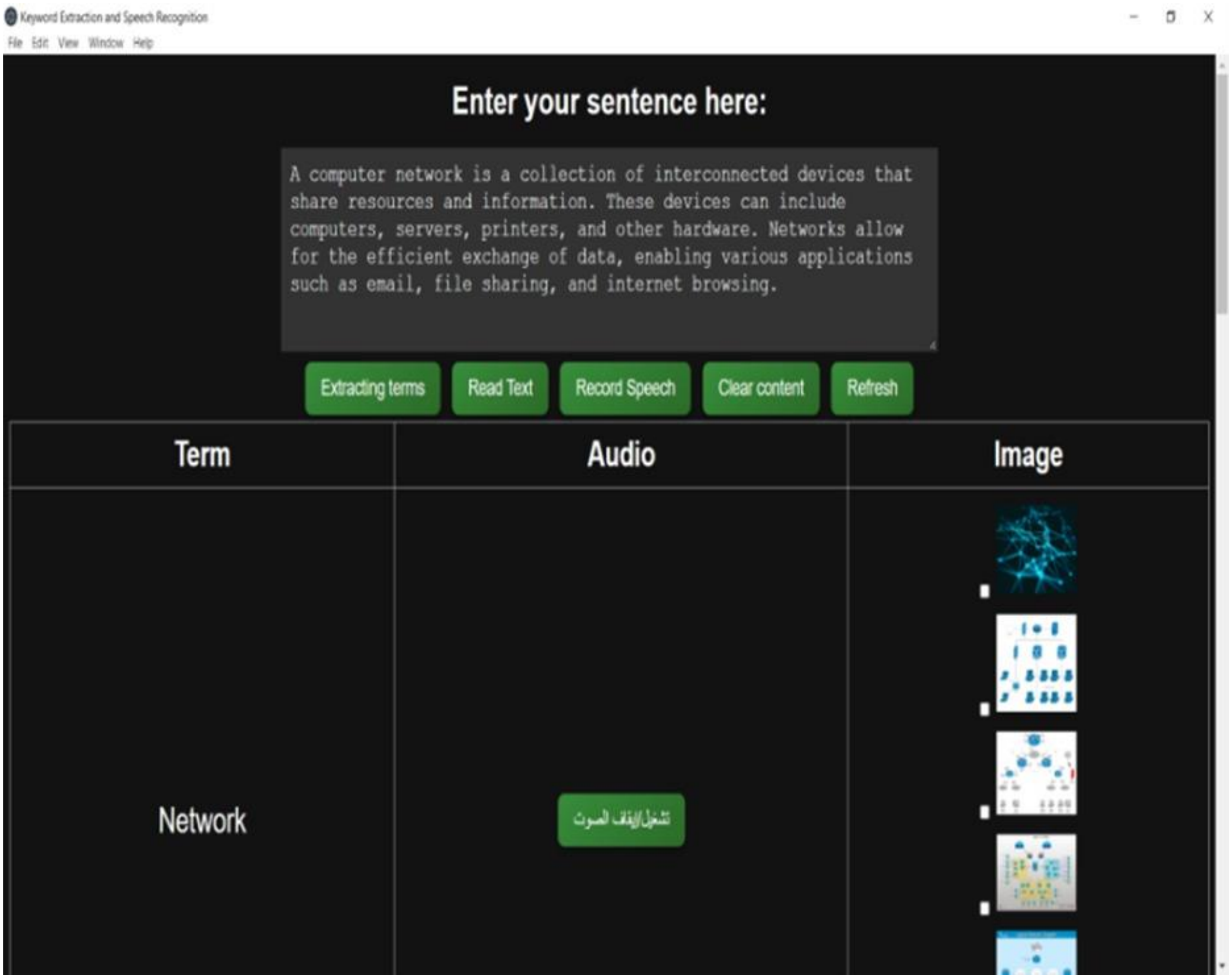

**Figure 3.** GUI to extract terminology from sentences

Figure 4 display the result of executing image integration after selecting the displayed images for each term from the image database and as shown in the format, the video can be watched without sound in the system. In the next stage, the audio is combined with the video by pressing the audio combination button opens the window in which the audio files for each term are displayed so that the audio files are combined and then the video without audio is combined with the combined audio file.

After audio file mashups selected from the private audio database for each extracted term of the mashup technique based on FFmpeg, an external tool controlled by Node.js code to merge video and audio files together without recoding the video

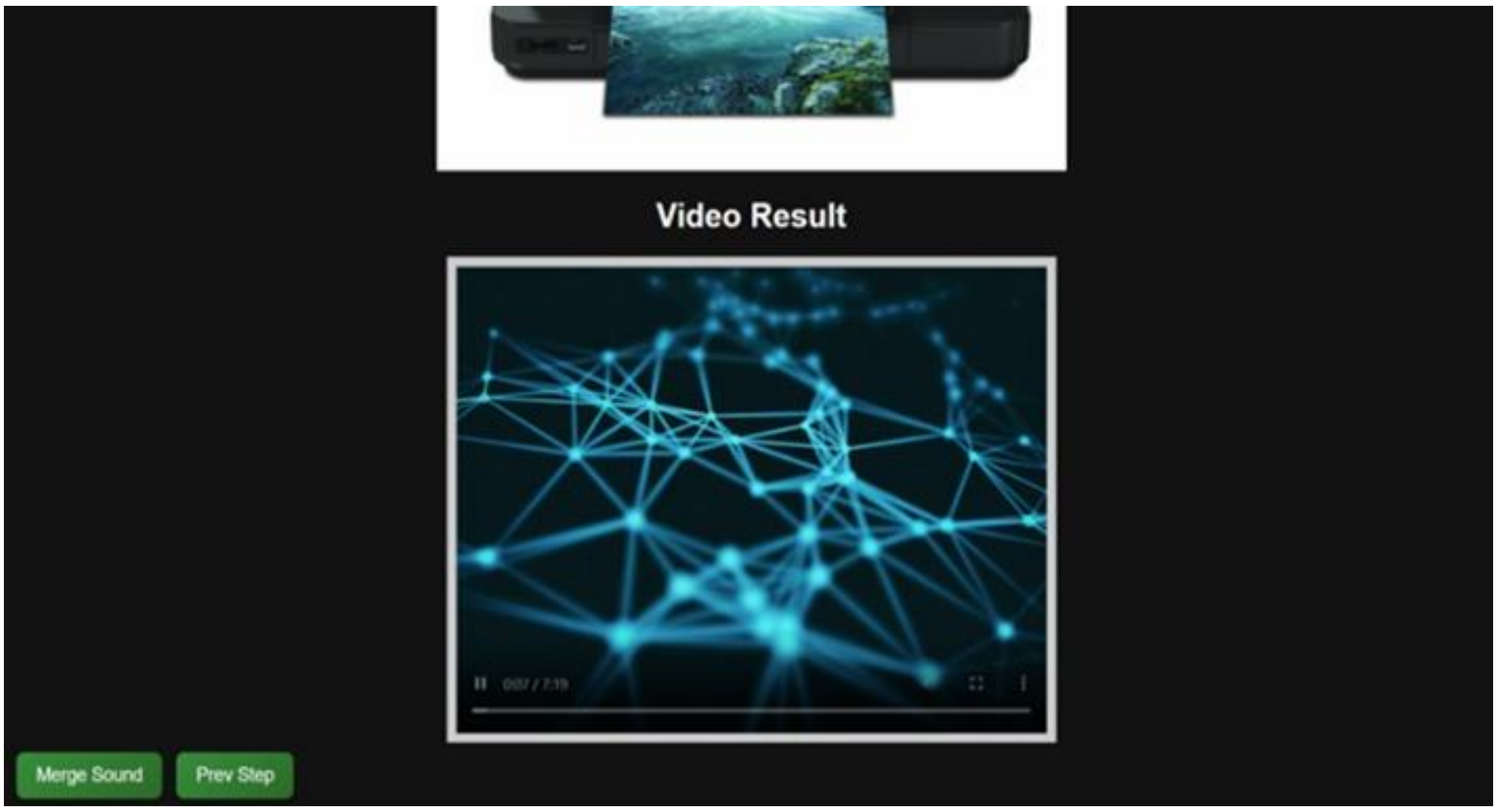

**Figure 4.** GUI displays the resulting video through TiVGAN technology.

The final form of the video reflects some of the integration of the sound-free video with the audio file through the FFMPEG library. Through this screen, the video can be seen in the system and can be downloaded on the user's computer as shown in figure (5).

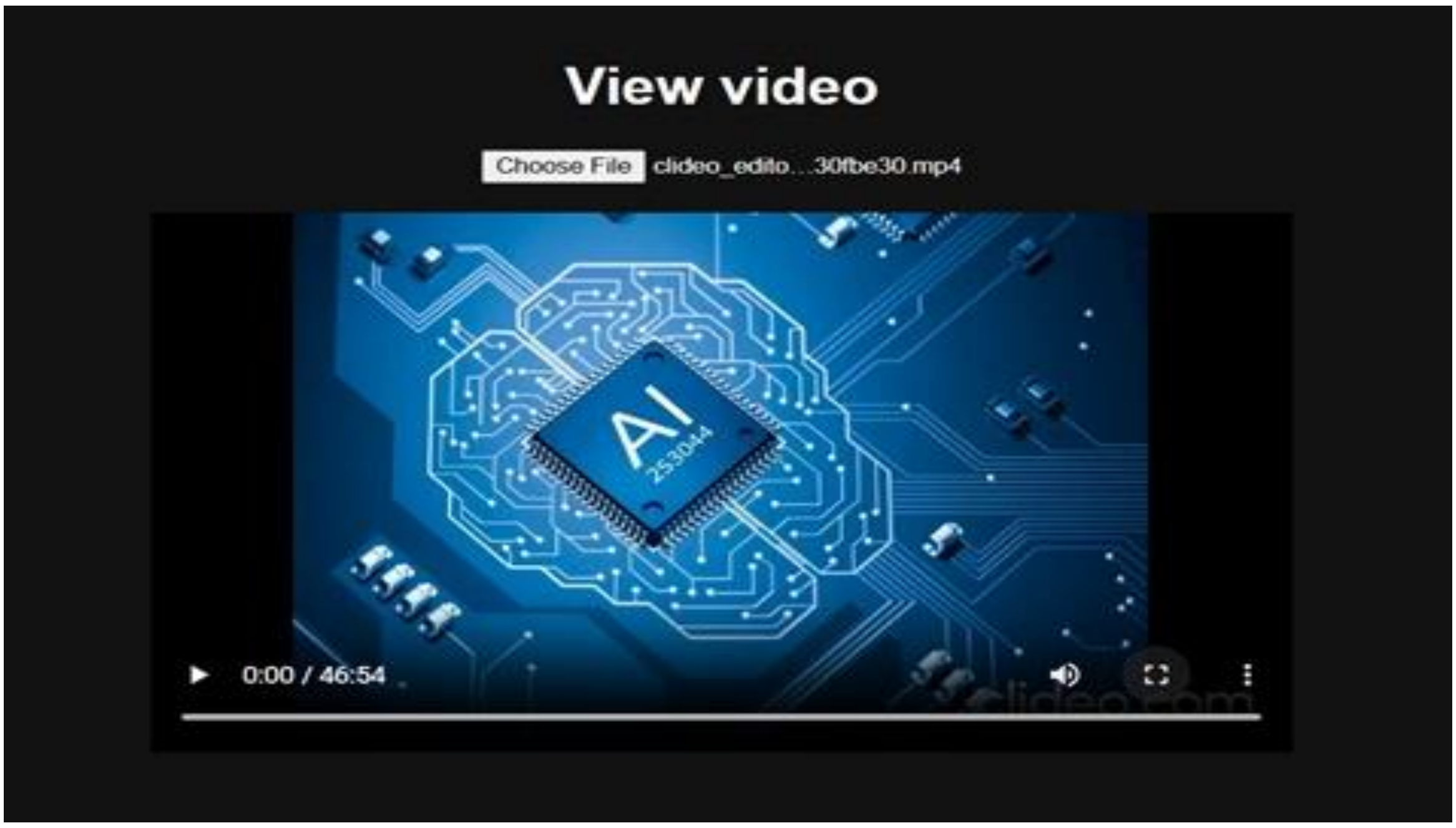

**Figure 5.** GUI showing the final video created by combining silent visuals with audio using TiVGAN.

**Quantitative results:**

For a quantitative evaluation, Frecht insection distance (fid) is used to measure the quality of the image generated, as shown in equation 4. FID is measured as a similarity between two image sets. Six video frames are collected from each method, and then FID is calculated between the same number of video frames in each image set and the training dataset. The best performance is shown by results, as presented in Table 2.

$FID = ||\mu_r - \mu g||^2 + Tr(\Sigma_r + \Sigma g - 2(\Sigma_r \Sigma g)^{\wedge}½)$ (4)

**Where:**

- $\mu_r$: Mean vector of the features extracted from the real images.
- μg: Mean vector of the features extracted from the generated images.
- $\Sigma_r$: Covariance matrix of the features from the real images.
- Σg: Covariance matrix of the features from the generated images.
- Tr: Trace operator, which sums the diagonal elements of a matrix.
- $(\Sigma_r \Sigma g)^{1/2}$: Matrix square root of the product of $\Sigma_r$ and Σg.

To calculate the Frechet Insection Distance (FID), the features are first extracted from images or video frames using a pretered incense V3 network. Then, both real and generated image distribution are calculated through both real and generated image distribution. Finally, the fréchet distance is calculated using the standard FID formula between these two distributions, which provides a quantitative measurement of equality between real images in terms of convenience distribution.

**Table 2:** Evaluate FID (Fréchet Inception Distance) for a range of models

| | **Without Text or Sound** | **with the text** | **With text and sound** |
|---|---|---|---|
| **TGAN [13]** | 70.10 | 59.59 | 55.30 |
| **MoCoGAN [14]** | 83.07 | 81.85 | 78.45 |
| **TGANS-C [15]** | 75.50 | 69.92 | 65.80 |
| **The Proposed System** | 40.25 | 35.10 | 28.75 |

The Proposed System is clearly the best, having the lowest FID in all cases, especially when using text and sound. The less the FID, the better the quality of the videos generated is closer to reality.

Sound inclusion benefits most models, but in the case of the Proposed System, this becomes an advantage that makes it remarkably the best. MoCoGAN uses sound less since its model poorly benefits from additional information. Proposed System outshines when using text and sound altogether, meaning it uses additional information to upgrade video quality properly.

To evaluate the effectiveness of the proposed system, a comparison was conducted against four state-of-the-art video generation models: TGAN, MoCoGAN, TGANS-C, and the Proposed System. The evaluation was based on a quantitative performance metric (e.g., FID score or user satisfaction rate), under three different input configurations Without text or sound, With text only, With both text and sound.

The results show that the proposed system significantly outperforms the other models across all three input conditions. The lower the score, the better the quality, which indicates that the proposed system produces more accurate, realistic, and semantically aligned educational videos.

TGAN and MoCoGAN demonstrate strong performance in generic video generation, particularly without specific input guidance (text or sound). However, their performance degrades when tasked with aligning to semantic content due to the absence of deep contextual understanding. TGANS-C incorporates semantic conditioning and shows improved alignment with text, but still lacks multimodal integration, which limits its applicability for rich educational content. Proposed System (though incomplete in the current evaluation) is designed for dual-modality input (text + image), but it does not fully leverage sound cues, making it less suitable for narrated educational video generation.

In contrast, the proposed system integrates NLP, TTS, and GANs in a unified pipeline, enabling it to extract semantic meaning from text, generate synchronized sound narration, and visualize concepts in a pedagogically meaningful way. This synergy leads to better alignment between the content and the visuals, especially when both text and sound are provided.

The proposed system emerges as the best performer in the context of educational video synthesis, particularly when high accuracy, relevance, and semantic depth are critical. Its ability to adapt to both textual and auditory input makes it superior to models like TGAN and MoCoGAN, which are not optimized for structured educational use cases.

When the proposed system was presented to experts, a high level of agreement was expressed by a large percentage of them, with 45% indicating "Strongly Agree" and 30% indicating "Agree." A smaller group (15%) recorded neutrality, while opposition was expressed by only a minority, with 5% selecting "Disagree" and another 5% selecting "Strongly Disagree." Overall, a generally positive perception of the program was conveyed, though complete consensus was not achieved as shown in Figure 7.

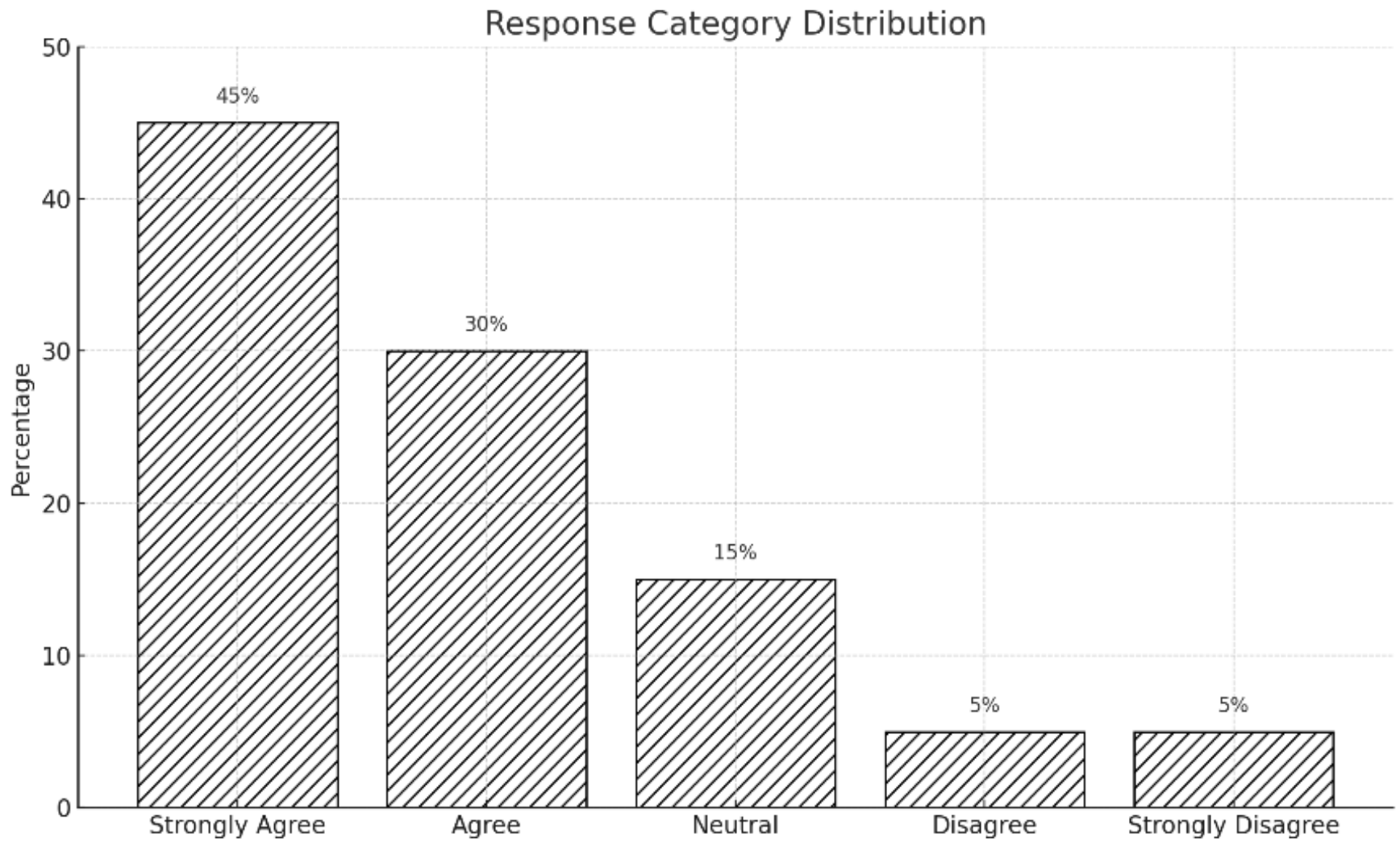

**Figure 7.** Chart showing the distribution of responses from experts in the program's questionnaire

A study was conducted with 15 faculty experts in computer science and educational technology. This aims to evaluate the computer -based system to generate instructional videos. The reactions were divided into five levels: strongly agreed, agreed, neutral, disagree and strongly disagree. Table (3) shows the items of the expert evaluation questionnaire for the proposed system according to 14 targeted criteria.

**Table 3:** Shows the items of the expert evaluation questionnaire for the proposed system

| No | Evaluation Item | Strongly Agree | Agree | Neutral | Disagree | Strongly Disagree |
|---|---|---|---|---|---|---|
| 1 | Assists computer science trainees in generating instructional videos. | | | | | |
| 2 | Enables user control over content sequencing and navigation. | | | | | |
| 3 | Presents video content in a logical, organized manner. | | | | | |
| 4 | Offers a user-friendly interface and intuitive interaction. | | | | | |
| 5 | Achieves its intended educational outcomes. | | | | | |
| 6 | Handles user input errors gracefully. | | | | | |
| 7 | Operates reliably without technical failures. | | | | | |
| 8 | Compatible with various computing systems. | | | | | |
| 9 | Avoids unnecessary visual clutter in videos. | | | | | |
| 10 | Makes optimal use of screen space. | | | | | |
| 11 | Delivers clear audio quality. | | | | | |
| 12 | Provides appropriate system response time. | | | | | |
| 13 | Matches video duration with learning content. | | | | | |
| 14 | Clearly communicates system goals to the user. | | | | | |

The 75% agreement fee highlights the gadget's robust educational design and technical reliability. Experts cited numerous strengths. It helps learner independence (Items 1, 2). It guarantees clear instruction and defined desires (Items 3, five, 14). It additionally suggests technical stability (Items 7, eight, 12).Only 10% of responses pointed to regions wanting improvement. These have been minor and did now not affect the general effectiveness. The effects support using the device in trainer schooling packages to sell technology-based learning.

## 5. Conclusion

This paper presents a novel method for generating instructional videos from text or audio input. The system first extracts key terms from the input, and then generates images corresponding to each term to create a silent video. Finally, it adds relevant audio from a database to produce a high-quality educational video. The proposed system incorporates GANs, ASR, NLP, and additional algorithms, with audio integration as a key distinguishing feature. Experts evaluated the system positively. It achieved the lowest error rate (28.75%) compared to other models, demonstrating superior performance in combining text and audio. Competing systems showed less improvement and failed to achieve comparable results. Future work may involve extending the system's application to other fields.

## References

[1] M. S. Noorderwier and M. van der Schoot, "The effectiveness of educational videos: A meta-analytic review of the literature," *Educational Research Review*, vol. 39, p. 100522, 2023.

[2] K. Kavitha et al., "The Transformative Trajectory of Artificial Intelligence in Education: A Bibliometric Analysis," *Journal of Educational Computing Research*, 2024.

[3] M. Mohammadi et al., "Artificial Intelligence in Multimodal Learning Analytics: A Systematic Literature Review," *Computers and Education: Artificial Intelligence*, 2025.

[4] A. Bewersdorff et al., "Taking the next step with generative artificial intelligence: The transformative role of multimodal large language models in education," *Journal of Computer Assisted Learning*, 2025.

[5] Q. Zhang and L. Chen, "Exploring the potential of generative AI for educational video creation: opportunities, challenges, and future directions," *Education and Information Technologies*, 2024.

[6] Y. Wang and Y. Zhang, "AI-powered tools for automated educational content creation: A systematic review," *Computers & Education*, vol. 104815, 2024.

[7] K. Yan, Y. Lin, and Y. Qiao, "Neural Video Representation for Continuous Video Generation," in *Proc. IEEE/CVF Conf. on Computer Vision and Pattern Recognition (CVPR)*, 2024.

[8] J. Zhangjie Wu, Y. Li, and B. Zhou, "T2VScore: A Reliable Metric for Text-to-Video Generation Evaluation," in *Proc. IEEE/CVF Conf. on Computer Vision and Pattern Recognition (CVPR)*, 2024.

[9] M. Mao et al., "TAVG: Text-to-Audio-Visual Generation with 1.7M Video Dataset and Contrastive Latent Diffusion," arXiv preprint arXiv: 2403.00123, 2024.

[10] Y. Ma et al., "Pose-Guided Text-to-Video Generation using Pose-Free Videos," in *Proc. International Conf. on Computer Vision (ICCV)*, 2023.

[11] Z. Zhu et al., "Text-Image-to-Video Generation: Controllable Video Synthesis from Static Images and Text Descriptions," in *Proc. IEEE/CVF International Conf. on Computer Vision (ICCV)*, 2023.

[12] A. Singer et al., "Make-A-Video: Text-to-Video Generation without Text-Video Data," arXiv preprint arXiv: 2209.14792, 2022.

[13] M. Saito and S. Saito, "TGANv2: Efficient Training of Large Models for Video Generation with Multiple Subsampling Layers," arXiv preprint arXiv: 1811.09245, 2020.

[14] Y. Tian et al., "A Good Image Generator Is What You Need for High-Resolution Video Synthesis," arXiv preprint arXiv: 2104.15069, 2021.

[15] D. Kim, D. Joo, and J. Kim, "TiVGAN: Text to Image to Video Generation with Step-by-Step Evolutionary Generator," arXiv preprint arXiv: 2009.02018, 2020.